# Local ferroelectricity in SrTiO$_3$ thin films

Oleg Tikhomirov

*Department of Physics and Astronomy, University of Pittsburgh,*
*3941 O'Hara Street, Pittsburgh, PA 15260*
*Institute of Solid State Physics, Chernogolovka, Russia 142432*

Hua Jiang

*Corning Applied Technologies, Woburn, MA 01801*

Jeremy Levy

*Department of Physics and Astronomy, University of Pittsburgh,*
*3941 O'Hara Street, Pittsburgh, PA 15260*

The temperature-dependent polarization of SrTiO$_3$ thin films is investigated using confocal scanning optical microscopy. A homogeneous out-of-plane and inhomogeneous in-plane ferroelectric phase are identified from images of the linear electrooptic response. Both hysteretic and non-hysteretic behavior are observed under a dc bias field. Unlike classical transitions in bulk ferroelectrics, local ferroelectricity is observed at temperatures far above the dielectric permittivity maximum. The results demonstrate the utility of local probe experiments in understanding inhomogeneous ferroelectrics.

The anomalous properties of strontium titanate (SrTiO$_3$) have resulted in many investigations[1-7]. At high temperatures, the dielectric response exhibits a Curie-Weiss dependence typical for a paraelectric-ferroelectric phase transition, *i.e.*, $\varepsilon_1(T) \propto (T - T_C)^{-1}$, with $T_C \approx 35$ K [4]. However, instead of decreasing below $T_C$, the permittivity saturates at a high value $\varepsilon_1(0) \approx 24000$. This behavior is observed in both pure single crystals [2, 3, 8] and ceramics [9]. The suppression of a ferroelectric phase transition in SrTiO$_3$ has been attributed to large ground state quantum fluctuations of the soft mode, giving rise to the term "quantum paraelectric" [4, 5].

The properties of SrTiO$_3$ are extremely sensitive to dopants and external perturbations. Ferroelectricity can be induced as a result of small concentrations of dopants (Ca, Bi, La…) [10], isotopic substitution in the oxygen octahedra [11], applied electric fields [8] or mechanical pressure [12, 13]. The development of thin film technologies has provided additional ways of controlling structure-sensitive properties. Biaxial strain arising from interaction with a cubic substrate can act as a substitute for external field [14, 15]. Recent calculations predict that thin films of pure SrTiO$_3$, contrary to bulk materials, should exhibit various ferroelectric and antiferrodistortive phases, depending on the magnitude and sign of the misfit strain [16].

Experimental observation of phase transitions in SrTiO$_3$ thin films is complicated by several factors. Perturbations of cubic SrTiO$_3$ are expected to result in weak or "incipient" ferroelectricity, with narrow hysteresis loops [2]. Inhomogeneous strain arising from relaxation at the film/substrate interface and from other structural defects can make these weak signatures difficult to observe using traditional methods. In these circumstances, spatially resolved techniques are preferable because fluctuating quantities are not averaged out.

Here we report observations of local ferroelectricity in SrTiO$_3$ films, probed by confocal imaging of the linear electrooptic response. SrTiO$_3$ films are grown on LaAlO$_3$ substrates using a metal organic chemical liquid deposition (MOCLD) method. In the MOCLD process, strontium acetate (Sr(AcO)$_2$) and titanium diisopropyl bisacetylacetonate (Ti(O-iPr)$_2$(acac)$_2$) were used as precursors. Atomic force microscopic analysis indicated that the surface roughness of the film was around 20 Å. Interdigitated Au/Ti electrodes were fabricated on the top surface by photolithography and lift-off techniques. The thickness of the SrTiO$_3$ film is 800 nm, intermediate Ti layer is 20 nm thick, and electrode itself is 150 nm thick.

The electrooptic response of the SrTiO$_3$ is measured using confocal scanning optical microscopy [17, 18]. Light from a HeNe laser is passed through a spatial filter and focused to a diffraction-limited spot on the sample (spot diameter $d \approx 0.5\ \mu m$). A combined dc+ac electric field ($E(t) = E_{dc} + E_{ac}\cos(\Omega t)$, $\Omega/2\pi = 5$ kHz) is applied to interdigitated surface electrodes (gap spacing $L = 25 \mu m$) to induce an electrooptic modulation of the reflected light. The normalized reflected light is measured with a sensitive balanced photodetector. A lock-in amplifier records the in-phase ($\tilde{I}_{1,\Omega}$) and 90°-shifted ($\tilde{I}_{2,\Omega}$) components of the linear electrooptic response as a function of position (x,y) and bias field $E_{dc}$. Images of the electro-optic response [$\tilde{I}_{1,\Omega}(x,y)$, $\tilde{I}_{2,\Omega}(x,y)$] are formed by raster scanning the laser beam across the sample surface. The magnitude of the in-phase signal $|\tilde{I}_{1,\Omega}(x,y)|$ provides a direct signature of ferroelectricity at $(x,y)$.

Absence of a linear electrooptic effect ($|\tilde{I}_{1,\Omega}(x,y)| \rightarrow 0$) indicates either a high symmetry paraelectric phase or specific mutual orientation of ferroelectric axis, electric field, and light polarization canceling the resulting effect of applied electric field on the optical index [18].

Figure 1(a) shows the capacitance, proportional to the dielectric permittivity $\varepsilon_1$, as a function of temperature for the SrTiO$_3$ film and interdigitated structure. In contrast to (Ba,Sr)TiO$_3$ films [18], $\varepsilon_1(T)$ decreases monotonically over the entire temperature range examined (100 to 300K), in accordance with other dielectric investigations [19-21]. Figures 1(b-e) show CSOM images [$\tilde{I}_{1,\Omega}(x,y)$] obtained at various temperatures. The high-temperature response [down to $T$=180 K, Fig. 1(b)] shows $\tilde{I}_{1,\Omega}(x,y)=0$ over the entire field of view. Small variations are due to shot-noise-limited intensity fluctuations of the laser source, which increase at the metal electrodes due to the high reflectance. As the sample is cooled, areas with pronounced electrooptic response begin to appear. At $T$=164 K [Fig. 1(c)], $\tilde{I}_{1,\Omega}(x,y)$ is negligibly small in the central region, but is non-zero and uniform near the edges of the electrodes. At $T$=154 K [Fig. 1(d)], areas between the electrodes start to develop a linear electrooptic response indicative of in-plane ferroelectricity. These signatures grow more pronounced at $T$=146 K [Fig. 1(d)], showing spatial variations down to the diffraction limit. Distinct regions with $|\tilde{I}_{1,\Omega}(x,y)|=0$ are observed in some areas of the sample at all temperatures $T >$ 100 K.

The field-dependent electrooptic response shows a variety of polarization states in the SrTiO$_3$ film. Examples of observed loops formed by measuring $\tilde{I}_{1,\Omega}(E_{dc})$ at various locations $(x,y)_i$ are presented in Fig. 2(a-c). Both the shape and the amplitude of these local tuning loops vary significantly with location. We observe (a) linear behavior, (b) linear behavior with saturation at high dc bias, and (c) hysteretic curves with high signal at zero bias. In many locations $\tilde{I}_{1,\Omega}(E_{dc})=0$ is found for all $E_{dc}$, similar to the paraelectric response observed at room temperature $T$=290 K [Fig. 2(d)].

Ferroelectricity in thin SrTiO$_3$ films was predicted in Ref. [16], and experimental evidence has been reported in Refs. [27, 28]. Different phases, both ferroelectric and non-polar ones, are predicted depending on the substrate misfit parameter. For the SrTiO$_3$/ LaAlO$_3$ system, there are two competing effects. The lattice constant of SrTiO$_3$ is larger than LaAlO$_3$, so the film is expected to be subject to two-dimensional compression. However, thin films typically relax over a distance that is inversely proportional to the misfit parameter (~12 nm for LAO/STO); this relaxation produces misfit dislocations and other defects which can create large local stresses in the film and lead to a spatial distribution of transition-related parameters. A second effect concerns the difference in thermal expansion coefficients between the film and substrate during cooling. This difference tends to favor tensile expansion of the film. For the relatively thick film considered here (800 nm), we expect the residual compression and thermal mismatch to be comparable in magnitude, leading to an inhomogeneous spatial distribution of dielectric properties.

Observation of areas with linear electrooptic response is consistent with the prediction of ferroelectric ordering in SrTiO$_3$ films. However, the electrooptic images show that the size of ordered regions possessing electrooptic response is small, at or below the diffraction limit, and most likely due to local stresses. The theory developed in [16] predicts both in-plane and out-of-plane ferroelectric polarization, depending on the magnitude and sign of the in-plane misfit parameter. To identify the predicted phases with observed types of electrooptic response we must analyze the electrooptic equation [17, 18]. The electrooptic signal is proportional to the induced deformation of the optical ellipsoid:

$$\Delta\eta = r_{13}E_z(x^2 + y^2) + r_{33}E_z z^2 + r_{42}(E_y yz + E_x xz) \qquad (1)$$

Here $x$, $y$, $z$ are the coordinates for an intersection of the light polarization vector with the optical index ellipsoid, and $z$ coincides with the ferroelectric polarization at a given point. The large relatively field-insensitive values of $|\tilde{I}_{1,\Omega}|$ [Fig. 2(c)] are consistent with in-plane ferroelectric polarization, as depicted in Fig. 3(a). In general, $\tilde{I}_{1,\Omega}$ depends on the angle between the polarization and electric field.

The linear field dependence of $\tilde{I}_{1,\Omega}$ [Fig. 2(a)] can be understood by assuming that the ferroelectric polarization is primarily out-of-plane, as depicted in Fig. 3(b). An in-plane bias field $E_{dc}$ deforms the optical ellipsoid along the direction of $E_{dc}$ and creates a non-zero contribution with $r_{42}$ in (1). For small values of $E_{dc}$ the response is linear, analogous to superparaelectricity or incipient in-plane ferroelectricity. Occasionally, saturated curves are observed [Fig. 2(b)], and are ascribed to completed switching into an in-plane polarization state, or saturation of the polarization. At zero bias, the $r_{42}$ contribution disappears, and the out-of-plane ferroelectricity no longer contributes to the linear electrooptic response, similar to a proper paraelectric [Fig. 3(c)]. Enhanced signals close to the electrodes [Fig. 1(c)] confirm that the out-of-plane polarization is non-zero. At the electrodes, the perpendicular component of the electric field couples directly to the out-of-plane polarization *via* $r_{33}$ (1). The alternating direction at each electrode results in alternating contrast, providing direct evidence that the polarization is uniform in direction.

It is remarkable that the observations of local ferroelectricity in SrTiO$_3$ films are not accompanied by corresponding signatures in $\varepsilon_1(T)$. Coexisting regions of paraelectricity and ferroelectricity are observed at temperatures far above the dielectric permittivity maximum. We believe this behavior may help explain some of the peculiar properties of bulk SrTiO$_3$. From permittivity measurements, this material is believed to be quantum paraelectric obeying Barrett law [1]; however, calculations show that quantum fluctuations are not strong enough to prevent ferroelectric displacement completely [4]. Moreover, narrow ferroelectric loops have been reported in pure SrTiO$_3$ samples at low temperature [2, 6, 11]. One possible explanation is that SrTiO$_3$ below *T*=35 K is really an incipient ferroelectric with non-zero net polarization, but its dielectric permittivity shows no "classical" decrease below this transition temperature due to a coexistence with virtually paraelectric quantum fluctuations, so the transition is not seen with conventional permittivity measurements.

In summary, we have performed electrooptic imaging of SrTiO$_3$ thin films. The linear electrooptic response confirms the existence of ferroelectricity at low temperature. Spatial distribution of the ferroelectric phase is likely driven by inhomogeneous stress. Local field-dependent experiments show both hysteretic and non-hysteretic behavior. It was shown that ferroelectricity in SrTiO$_3$ thin films is not accompanied by a corresponding signature in the dielectric permittivity, as it is in bulk materials.

The work was supported by the Office of Naval Research (N00173-98-1-G011) and the National Science Foundation (DMR-9701725).

Figure Captions

FIG. 1. Phase transition in SrTiO$_3$ thin film. (a) Capacitance *vs.* temperature. Circles indicate the temperatures where electrooptic CSOM images (b-e) were obtained: (b) 181K; (c) 164K; (d) 154K; (e) 146K.

FIG. 2. Examples of local CSOM tuning loops in SrTiO$_3$ thin film. (a) to (c) are measured from different places in the sample at *T*=100K; (d) *T*=290K.

FIG. 3. Possible orientations of local ferroelectric polarization and expected contributions to CSOM response. (a) – in-plane polarization, (b) – out-of-plane polarization, (c) – paraelectric state.

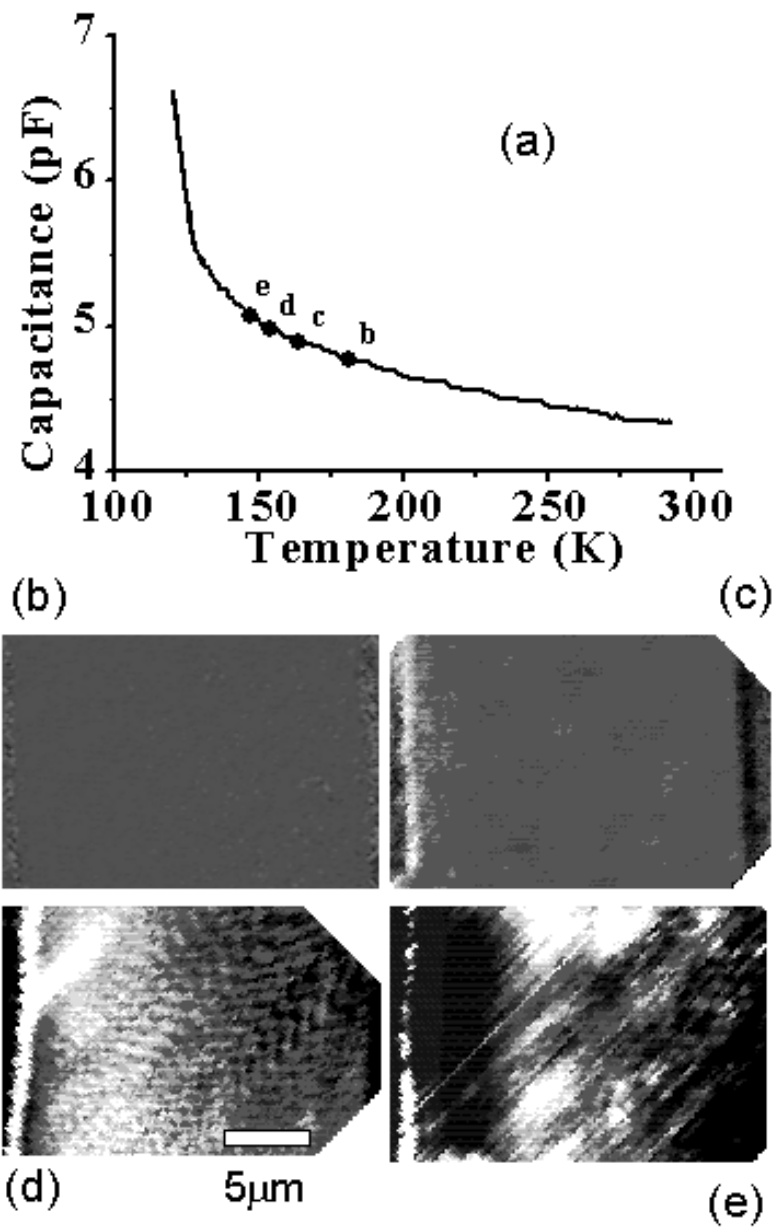

Figure 1

O. Tikhomirov, H. Jiang, and J. Levy

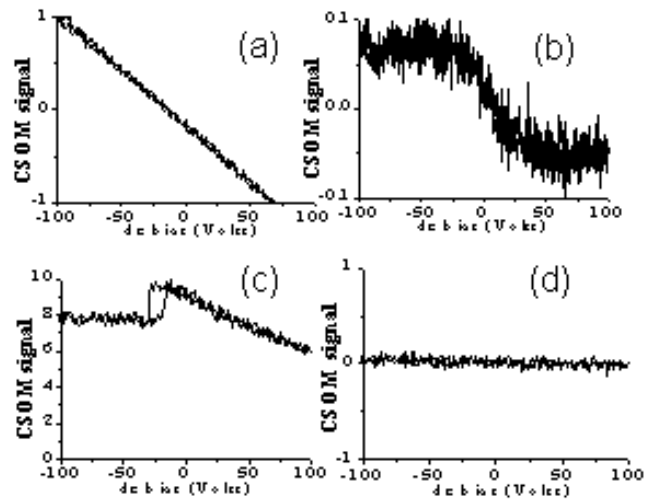

Figure 2

O. Tikhomirov, H. Jiang, and J. Levy

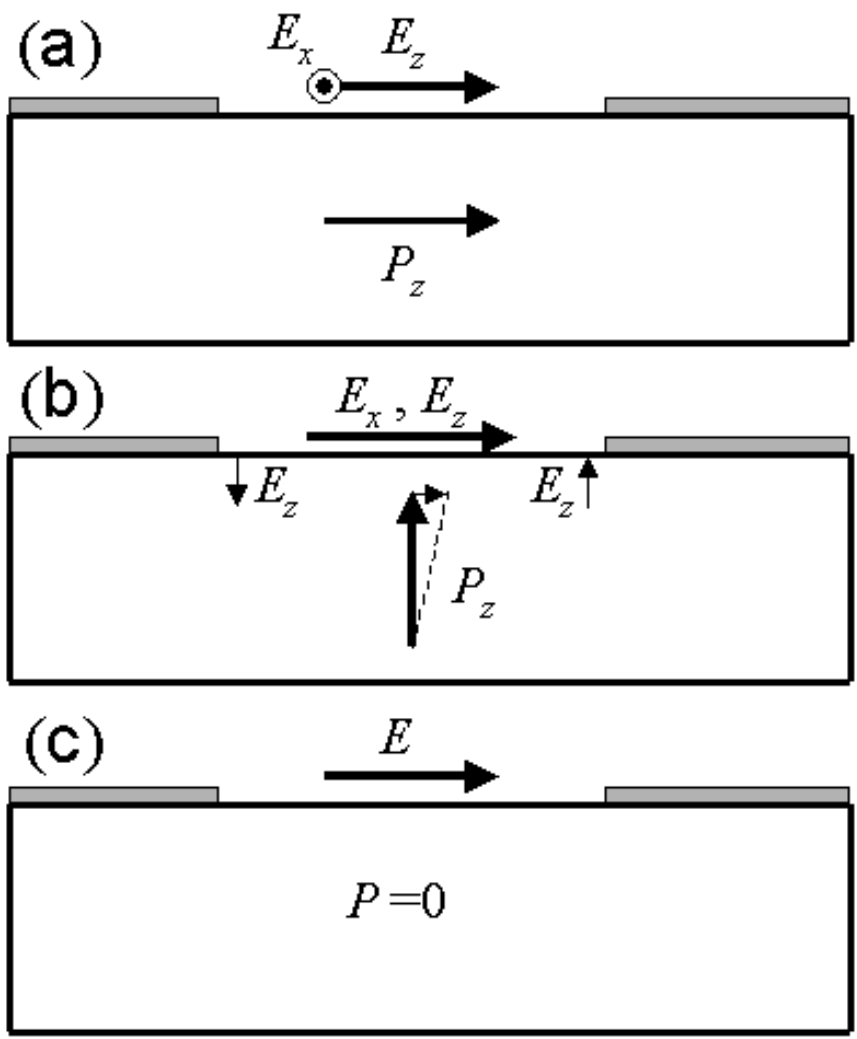

Figure 3

O. Tikhomirov, H. Jiang, and J. Levy